\documentstyle[12pt]{article}
\begin{document}
\begin{titlepage}
\baselineskip=0.30in
\begin{flushright}
November 1997
\end{flushright}
\vspace{0.2cm}

\begin{center}
{\Large Pion and Eta Strings  }
\vspace{.2cm}

\begin{center} Xinmin Zhang$^1$, Tao Huang$^1$ and Robert H. 
Brandenberger$^2$
 \end{center}

\vspace{.2in}
\it{
 $^1$Institute of High Energy Physics, Academia Sinica,
Beijing 100039, China}
\\
 
\it{ $^2$Department of Physics, Brown University, Providence, IR 02912, USA}

\end{center}

\vspace{.2cm}

\begin{footnotesize}
\begin{center}\begin{minipage}{6in}
\baselineskip=0.25in

\begin{center} ABSTRACT\end{center}

In this paper we construct a string-like classical solution, the pion-string,
in the linear sigma model. We then study the stability of the pion-string, and
find that it is unstable in the parameter space allowed experimentally.
We also speculate on the existance of an unstable eta-string, associated
with spontaneous breakdown of the anomalous $U_A(1)$ symmetry in QCD
at high temperatures. The implications of the pion and eta strings for cosmology
and heavy ion collisions are briefly mentioned.

\end{minipage}\end{center}
\end{footnotesize}
\vspace{.2cm}

\vskip 0.8cm \noindent BROWN-HET-1103 
\end{titlepage}
\eject

\baselineskip=0.25in

Strings, as classical solutions in theories with spontaneously broken 
symmetries, play an important role in both particle physics and cosmology. 
Thus it is of crucial importance to know if strings can exist 
in realistic models of
strong and electroweak interactions. Recently, in an inspiring paper,
Vachaspati[1] showed that string-like structures exist in the standard 
model of
electroweak theory. In this paper, we will consider strings in models
with spontaneously broken chiral symmetry of QCD. First of all, we
explicitly construct a classical solution, the pion string, in 
the linear sigma model, and then argue for the existence of an 
eta string at high tempratures.

In recent years, the
$SU_L(2) \times SU_R(2) \sim O(4)$ linear sigma-model has been
 often used as a model of hadron dynamics for chiral symmetry breaking in QCD, 
in particular in studies of the physics associated with the 
chiral phase transition[2] and of the disoriented chiral condensates[3] 
in heavy ion collisions. The lagrangian density of this model is
\begin{equation}
{\cal L} = \frac{1}{2} {( \partial_\mu \sigma )}^2 + \frac{1}{2} {( \partial_\mu
 {\vec\pi} )}^2 - \frac{\lambda}{4} ( \sigma^2 + {\vec{\pi}}^2 - f_\pi^2 ) + 
H \sigma .
\end{equation}
In this paper, for simplicity, we consider the chiral limit, $H=0$. 
In our discussion of the pion string it proves convenient
 to define the new fields  
\begin{eqnarray}
 \phi&=& \frac{ \sigma + i \pi^0}{ \sqrt 2 },\\
 \pi^\pm &=& \frac{ \pi^1 \pm i \pi^2 }{ \sqrt 2}.
\end{eqnarray}
The linear sigma model in eq.(1) now can be rewritten as
\begin{equation}
{\cal L} =
{(\partial_\mu \phi )}^* \partial^\mu \phi
+ \partial_\mu \pi^+ \partial^\mu \pi^- - \lambda {( \pi^+ \pi^-
+ \phi^* \phi - \frac{f_\pi^2}{2} )}^2.
\end{equation}
During chiral symmetry breaking, the field $\phi$ takes on a nonvanishing 
vacuum expectation value, which breaks $SU_L(2) \times SU_R(2)$ 
down to $SU_{L+R}(2)$. This results in a massive sigma and
three massless Goldstone bosons. In addition, we will demonstrate below that
there is a static unstable string-like solution, the pion-string.

For static configurations in eq.(4), the energy functional is given by
\begin{equation}
E = \int d^3 x 
\left [ {\vec{\bigtriangledown} \phi }^* \vec{\bigtriangledown}\phi
+ \vec{\bigtriangledown}\pi^+ \vec{\bigtriangledown}\pi^- + \lambda
{( \pi^+ \pi^- + \phi^* \phi - \frac{ f_\pi^2}{2} )}^2 \right ].
\end{equation}
The time independent equations of motion are:
\begin{eqnarray}
 {\bigtriangledown}^2 \phi &=&  2 \lambda ( \pi^+ \pi^-
+ \phi^* \phi - \frac{f_\pi^2}{2} ) \phi,\\
 \bigtriangledown^2 \pi^+&=& 2 \lambda ( \pi^+ \pi^- + \phi^* \phi
   - \frac{f_\pi^2}{2} ) \pi^+ .
\end{eqnarray}
The pion string solution extremising the energy functional in eq.(5) is
given by
\begin{eqnarray}
\phi &=& \frac{ f_\pi}{\sqrt 2} \rho (r) e^{i n \theta}, \\
\pi^\pm &=& 0,
\end{eqnarray}
where the coordinates $r$ and
$\theta$ are polar coordinates in the $x-y$ plane, and the integer $n$ is the winding number. In the following discusion, we will restrict ourselves to $n=1$.

Substituting (8) into (6), we obtain the equation of motion for $\rho(r)$,
\begin{equation}
\frac{1}{r} \frac{\partial}{\partial r}
( r \frac{\partial}{\partial r} )
\rho(r)
- \frac{1}{ r^2 } \rho(r)
= \lambda f_\pi^2 ( \rho^2 - 1 ) \rho (r) .
\end{equation}
The boundary conditions for $\rho(r)$ are
\begin{eqnarray}
r \rightarrow 0, & & \rho(r) \rightarrow 0 ; \\
r \rightarrow \infty, & & \rho(r) \rightarrow 1.
\end{eqnarray}

In principle, $\rho(r)$ can be determined numerically by solving 
Eq.(10) with boundary conditions in (11) and (12). But it is simpler to adopt
a variational approach. Following 
the method of Hill, Hodges and Turner[5], we
 make an ansatz of the following form:
\begin{equation}
\rho(r) = ( 1 - e^{- \mu r} ),
\end{equation}
where $\mu$ is a variational parameter.

Adopting expression (13) as a variational ansatz, it follows
 that the energy per unit length of the string is
\begin{equation}
E_{(pion-string)}=
\frac{1}{4} \pi f_\pi^2
+ \pi f_\pi^2 I_{\theta}( \mu, R) + \frac{\pi}{2} \frac{89}{144}
\lambda f_\pi^2 \frac{f_\pi^2}{\mu^2} ,
\end{equation}
where $ I_{\theta}( \mu, R)
= \int_0^\infty \frac{dr}{r} {( 1 - e^{- \mu r} )}^2 \approx \ln (\mu R)$
is logarithmically divergent, as expected 
in a theory with global symmetry breaking. Thus, we introduce a
large-scale cutoff $R$. However, the $R$ dependence will disappear
upon varying with respect to $\mu$:
\begin{equation}
\frac{\partial I_{\theta}(\mu, R)}{\partial \mu} |_{R\rightarrow \infty}
= \frac{1}{\mu}.
\end{equation}

The solution for $\mu$ is obtained by varying the
 energy functional of the pion-string in (14) with respect to $\mu$:
\begin{equation}
\mu^2 = \lambda \frac{89}{144} f_\pi^2.
\end{equation}
Substituting (16) into (14), the mass of the pion-string per unit length
is
\begin{equation}
E_{(pion-string)} = \left [
 \frac{3}{4} + I_{\theta}(\mu, R) \right ] \pi f_\pi^2
     \approx \left [ \frac{3}{4} + \ln (\mu R) \right ] \pi f_\pi^2 .
\end{equation}

The pion-string is not topologically stable, since any field configuration 
can be continuously deformed to the vacuum. To study the stability
of the pion-string, we consider infinitesimal perturbations of
the field $\pi^\pm$ and check if the variation in the energy is positive 
or negative.

Discarding terms of cubic and higher orders in $\pi^\pm$, we find
\begin{equation}
E = E_{(pion-string)} + \delta E ,
\end{equation}
where
\begin{equation}
\delta E = \int d^3 x \left [
\vec{\bigtriangledown} \pi^+ \vec{\bigtriangledown} \pi^-
+  \lambda f_\pi^2 ( \rho^2 - 1 ) \pi^+ \pi^- \right ].
\end{equation}
Following Ref.[4], we consider an expansion of the $\pi^\pm$ fields
in Fourier modes,
\begin{equation}
\pi^+ = \chi_m(r) e^{im \theta} .
\end{equation}
Inserting the expressions for the m-th mode of $\pi^\pm$ in eq.(19) gives
\begin{equation}
\delta E
= \int 2 \pi r dr \left [
{ ( \frac{\partial \chi_m}{\partial r} )}^2 +
  \frac{m^2}{r^2} \chi_m^2 + \lambda f_\pi^2 ( \rho^2 - 1 ) \chi_m^2
           \right ],
\end{equation}
where the first term (the kinetic energy part) and the second term are always
positive, but the third term, the potential energy, is negative. Notice that the
second term gives the smallest contribution to the positive energy in $\delta E$
for $m=0$, so we will focus on $m=0$.

Defining $\xi = f_\pi r, \chi = f_\pi R$, 
and setting $m=0$, $\delta E$ becomes:
\begin{equation}
\delta E = 2 \pi f_\pi^2 \int \xi d\xi \left [
{( \frac{\partial R}{\partial \xi} )}^2 + \lambda ( \rho^2 - 1) R^2 \right ].
\end{equation}
After an algebraic computation we can rewrite $\delta E$ in the form
\begin{equation}
\delta E = 2 \pi f_\pi^2 \int \xi d\xi R \hat{O} R,
\end{equation}
where
\begin{equation}
\hat{O} = - \frac{1}{\xi} \frac{\partial}{\partial \xi}
           ( \xi \frac{\partial}{\partial \xi} ) + \lambda ( \rho^2 - 1) .
\end{equation}
The question of stability of the pion-string reduces to checking if the 
eigenvalues of the operator $\hat{O}$ in its spectrum are negative,
 subject to the eigenfunction $R$ satisfying the boundary conditions $R( \xi
\rightarrow 0) \rightarrow $  constant,
and $R( \xi \rightarrow \infty ) \rightarrow 0$.

To simplify the analysis, we take a variational approach, making
use of an ansatz of the form[5],
\begin{equation}
R= \sigma_0 e^{- \kappa \xi} ( 1 + \kappa \xi + \kappa_1 \xi^2 +
 \kappa_2 \xi^3 ),
\end{equation}
where $\sigma_0, \kappa, \kappa_1, \kappa_2$ are dimensionless variational 
parameters. This ansatz has the correct short-distance limit and $\kappa^{-1}$
represents the size of the $\pi^\pm$ condensate on the pion-string. By inserting (25) into (23) it is obvious that the negative term wins out if 
$\lambda \geq 1$. This implies that
the pion-string is unstable in the parameter space allowed experimentally 
( $\lambda \sim 10 - 20$[2]).
It can be shown by numerical analysis that $\delta E$ is positive only
for very small values of $\lambda$ ($\lambda \leq 10^{-8}$), and
hence the pion-string is only stable for these values. 

In the early universe and in heavy-ion collisions, pion strings are expected to
be produced and to subsequently decay.
 Their lifetime can be estimated by considering their
 interactions with the surrounding plasma. Based on
 a naive dimensional analysis, their lifetime $\tau$ should
 be proportional to the inverse of the corresponding temperature. For strongly
interacting theory such as the linear sigma model studied here, we
expect that $\tau = O(1) T^{-1}$, where
$T$ is the temperature at the time when the chiral symmetry of QCD is restored.

Before concluding, let us speculate about
 the existence of an unstable $\eta$ string. 
In QCD, in the limit of massless quarks, there is an additional
 $U_A(1)$ chiral symmetry. This chiral symmetry, 
when broken by the quark condensate, predicts the existence of a goldstone
 boson. There is no such a light meson, however. This is resolved by
the Adler-Bell-Jackiw $U_A(1)$ anomaly together with the properties of
non-trival 
vacuum structure of non-abelian gauge theory, in particular QCD. The
$U_A(1)$ symmetry is badly broken by instanton effects at zero temperature.

As the density of matter and/or the temperature increases, it is expected 
that the instanton effects will rapidly disappear[6],
 and one thus has an additional $U_A(1)$ symmetry
 (besides $SU_L(2) \times SU_R(2)$)
at the transition temperature of the QCD chiral symmetry.
 When the $U_A(1)$ symmetry is broken spontaneously 
by the quark condensate, a topological string, the $\eta$-string, results.
Differing from the pion-string, the $\eta$-string is topologically stable at
high temperatures, but will decay as the temperature decreases.
The $\eta$-string can form during the chiral phase 
transition of QCD. In the setting of cosmology, it will
 exist during a specific epoch below the QCD 
chiral symmetry breaking temperature during the evolution of the 
universe. In the context of heavy-ion collisions,
 it will exist in the plasma created by the collision
 during a period shortly after the cooling 
of the interaction region below the
 symmetry breaking scale. 
The strings then become unstable as the temperature decreases and when
 the instanton effects become substantial.

\begin{center}{\Large  Acknowledgements}\end{center}

We thank Jin Hong-Ying and Cao Jiun-Jer for discussions and help in the
numerical calculation. This work 
was supported in part by National Natural Science Foundation of China and by the U.S. Department of Energy under Contract DE-FG0291ER40688.
\vspace{1cm}

{\LARGE References}
\vspace{0.3in}
\begin{itemize}
\item[{\rm[1]}]  
T. Vachapati, Phys. Rev. Lett. 68, 1977 (1992); Nucl. Phys. B397, 648 (1993).

\item[{\rm[2]}] 
For example, see, R.D. Pisarski, "Applications of Chiral Symmetry"
(hep-ph/9503330).

\item[{\rm[3]}] 
K. Rajagopal and F. Wilczek, Nucl. Phys. B379, 395 (1993);
B404, 577 (1993).

\item[{\rm[4]}]  
M. James, L. Perivolaropoulos, and T. Vachaspati, Nucl. Phys. B395, 534 (1993).

\item[{\rm[5]}]   
C.T. Hill, H.M. Hodges and M. Turner, Phys. Rev. D37, 263 (1988).

\item[{\rm[6]}]   R.D. Pisarski and F. Wilczek, Phys. Rev. D29, 338 (1984);
E. Shuryak, Comm. Nucl. Part. Phys. 21, 235 (1994), and refs therein;
J. Kapusta, D. Kharzeev and L. McLerran, Phys. Rev. D53, 5028 (1996).
\end{itemize}
\end{document}